\documentclass[%
 reprint,
showpacs,
nofootinbib,
 amsmath,amssymb,
 aps,
prd,
]{revtex4-1}

\usepackage{graphicx}
\usepackage{epstopdf}
\usepackage{dcolumn}
\usepackage{bm}

\begin{document}


\title{Dynamical analysis for a vector-like dark energy}

\author{Ricardo C. G. Landim}
\email{rlandim@if.usp.br}
 \affiliation{%
 Instituto de F\'isica, Universidade de S\~ao Paulo\\
 Caixa Postal 66318,  05314-970 S\~ao Paulo, S\~ao Paulo, Brazil
}%



\date{\today}

\begin{abstract}
In this paper we perform a dynamical analysis for a vector field as a candidate for the dark energy, in the presence of a barotropic fluid. The vector is one component of the so-called cosmic triad, which is a set of three identical copies of an abelian field pointing mutually in  orthogonal directions. In order to generalize the analysis, we also assumed the interaction between dark energy and the barotropic fluid, with a phenomenological coupling. Both matter and dark energy eras can be successfully described by the critical points, indicating that the dynamical system theory is a viable	tool to analyze asymptotic states of such cosmological models.


\end{abstract}

\maketitle



\section{Introduction}

Around $95\%$ percent of the universe today corresponds to two kinds of components whose nature is still unknown. The first one, called dark energy,  is believed to be responsible for the current accelerated expansion of the universe \cite{reiss1998, perlmutter1999} and it is  dominant at present times ($\sim$ 68\%) \cite{Planck2013cosmological}.  In addition to ordinary matter, the remaining $27\%$ of the energy content of the universe is a form of matter that interacts in principle only gravitationally, known as dark matter. The simplest dark energy candidate is the cosmological constant, whose equation of state $w_\Lambda=p_\Lambda/\rho_{\Lambda}=-1$ is in agreement with the Planck results \cite{Planck2013cosmological}. This attempt, however,  suffers from the so-called cosmological constant problem, a huge discrepancy of 120 orders of magnitude between the theoretical prediction and the observed data. Such a huge disparity motivates physicists to look into more sophisticated models. This can be done either looking for a deeper understanding of where the cosmological constant comes from, if one wants to derive it from first principles, or considering other possibilities for accelerated expansion. In the former case, an attempt is the famous KKLT model \cite{kklt2003}, and in the latter one, possibilities are even broader, with  modifications of General Relativity, additional matter fields and so on (see \cite{copeland2006dynamics, dvali2000, yin2005} and references therein).  Moreover, the theoretical origin of this constant is still an open question, with several attempts but with no definitive answer yet.

Among a wide range of alternatives, the field theory can provide some other candidates. The simplest one is the canonical scalar field   \cite{peebles1988,ratra1988,Frieman1992,Frieman1995,Caldwell:1997ii},  although non-canonical scalar fields have also been explored (tachyon field \cite{Padmanabhan:2002cp,Bagla:2002yn}, k-essence \cite{ArmendarizPicon:2000dh}, or supergravity-inspired models \cite{Brax1999,Copeland2000,Landim:2015upa,Landim:2015hqa}, for instance). Another dark energy candidate is a spin-1 particle, described by a vector field. To be consistent with the homogeneity and isotropy of the universe, there should be three identical copies of the vector field, which one with the same magnitude and pointing mutually in the orthogonal direction. They are called cosmic triad  and were proposed in  \cite{ArmendarizPicon:2004pm}. Other possibilities of vector dark energy are shown in \cite{Koivisto:2008xf,Bamba:2008ja,Emelyanov:2011ze,Emelyanov:2011wn,Emelyanov:2011kn,Kouwn:2015cdw}.

In addition, the two components of the dark sector may interact with each other \cite{Wetterich:1994bg,Amendola:1999er,Farrar:2003uw,Guo:2004vg,Cai:2004dk,Guo:2004xx,Bi:2004ns,Gumjudpai:2005ry,yin2005,Wang:2005jx,Wang:2005pk,Wang:2005ph,Wang:2007ak,micheletti2009,Costa:2014pba} (see \cite{Wang:2016lxa} for a recent review), since their densities are comparable and the interaction can eventually alleviate the coincidence problem \cite{Zimdahl:2001ar,Chimento:2003iea}. 

When the dark energy candidate is in the presence of a barotropic fluid (with an equation of state $w_m=p_m/\rho_m$)  the relevant evolution equations can be converted   into an autonomous system and the asymptotic states of the cosmological models can be analyzed. Such approach was done for uncoupled dark energy (quintessence, tachyon field and phantom field  for instance \cite{copeland1998,ng2001,Copeland:2004hq,Zhai2005,DeSantiago:2012nk,Dutta:2016bbs}) and coupled dark energy \cite{Amendola:1999er,Gumjudpai:2005ry,TsujikawaGeneral,amendola2006challenges,ChenPhantom,Landim:2015poa,Landim:2015uda,Mahata:2015lja}, but it remained to be done for a vector-like dark energy, whose interesting properties were explored in  \cite{ArmendarizPicon:2004pm}.  In this paper, we use  the linear dynamical systems theory to investigate the critical points that come from the evolution equations for the vector-like dark energy, considering also the possibility of interaction between the two components of the dark sector, where we propose a phenomenological coupling. The fixed points found can successfully describe  the matter-dominated universe and the current stage of accelerated expansion, provided that the interaction is sufficiently small.    

The rest of the paper is organized as follows. In section \ref{de} we present the basics of the interacting dark energy and the dynamical analysis theory. Section \ref{VDE} contains the dynamics of the vector-like dark energy in the light of the dynamical system theory, where the critical points and their stabilities are presented. Section \ref{conclu} is reserved for conclusions. We use Planck units ($\hbar=c=1 =M_{pl}=1$) throughout the text.

\section{Interacting dark energy and the dynamical system theory}\label{de}

As a generalization of the continuity equation, we  consider an interaction between the dark energy, described by the cosmic triad, and a barotropic fluid, in such a way that the total energy-momentum tensor is still conserved. Dark energy has an energy density $\rho_A$ and pressure $p_A$, with an equation of state  given by $w_A=p_A/\rho_A$. In the flat Friedmann--Lema\^itre--Robertson--Walker (FLRW) background with a scale factor $a\equiv a(t)$, the continuity equations are

\begin{equation}\label{contide}
\dot{\rho_A}+3H(\rho_A+p_A)=-\mathcal{Q},
\end{equation}

\begin{equation}\label{contimatter}
\dot{\rho_m}+3H(1+w_m)\rho_m=\mathcal{Q},
\end{equation}

\noindent respectively, where $H=\dot{a}/a$ is the Hubble rate,  $\mathcal{Q}$ is the coupling, and the dot is a derivative with respect to the cosmic time $t$. The index $m$ stands for the barotropic fluid, with $w_m=0$ for non-relativistic matter and $w_m=1/3$ for radiation. The case of $\mathcal{Q}>0$ corresponds to a dark energy transformation into the barotropic fluid, while $\mathcal{Q}<0$ is the transformation in the opposite direction. In principle, the coupling  can depend on several variables $\mathcal{Q}=\mathcal{Q}(\rho_m,\rho_A, \dots)$, so that, inspired by the quintessence case \cite{Wetterich:1994bg,Amendola:1999er}, where the coupling is $Q\rho_m \dot{\phi}$, we assume the phenomenological interaction $\mathcal{Q}=3 Q \rho_m\dot{A}/a$, where $Q$ is a positive constant. The coupling has this form in order for the right-hand side of the Proca-like equation (\ref{eqmotionvec}) to be no longer zero but to equal $Q\rho_m$.\footnote{In the scalar field case the coupling $Q\rho_m\dot{\phi}$ leads to the equation of motion in the FLRW background which also equals $Q\rho_m$.}   The case with negative $Q$ is similar and we will not consider it here because the minus sign of the case $\mathcal{Q}<0$ can be absorbed into $\dot{A}$, instead of considering $Q<0$.

To deal with the dynamics of the system, we define dimensionless variables. The new variables are going to characterize a system of differential equations in the form

\begin{equation}
X'=f[X],
\end{equation}

\noindent where $X$ is a column vector of dimensionless variables and the prime is the derivative  with respect to $ \log a$, where we set the present scale factor $a_0$ to be one. The critical points $X_c$ are those ones that satisfy $X'=0$. In order to study stability of the fixed points, we consider linear perturbations $U$ around them, thus $X=X_c+U$. At the critical point the perturbations $U$ satisfy the following equation:

\begin{equation}\label{Jacobian}
U'=\mathcal{J}U,
\end{equation}

\noindent where $\mathcal{J}$ is the Jacobian matrix. The stability around the fixed points depends on the nature of the eigenvalues ($\mu$) of $\mathcal{J}$, in such a way that they are stable points if they all have negative values, unstable points if they all have positive values and saddle points if at least one eigenvalue has positive (or negative) value, while the other ones have opposite sign.  In addition, if any eigenvalue is a complex number, the fixed point can be stable (Re $\mu<0$) or unstable (Re $\mu>0$) spiral, due to the oscillatory behavior of its imaginary part.

\section{Vector-like dark energy dynamics}\label{VDE}

The  Lagrangian for three identical copies of an abelian field (called cosmic triad in  \cite{ArmendarizPicon:2004pm}), here uncoupled to matter, is given by

\begin{equation}\label{LVDE}
 \mathcal{L}_A=-\sqrt{-g}\sum_{a=1}^3\left(\frac{1}{4}F^{a \mu\nu}F_{\mu\nu}^a+V(A^{a2})\right),
\end{equation} 

\noindent where $F_{\mu\nu}^a=\partial_\mu A^a_\nu -\partial_\nu A_\mu^a$ and $V(A^2)$ is the potential for the vector field, which breaks gauge invariance, with $A^{a2}\equiv A_\mu^a A^{a\mu}$. The energy-momentum tensor of the field is obtained varying the Lagrangian (\ref{LVDE}) with respect to the metric and it is $T_{\mu\nu}^A=\sum_{a=1}^3T^{a}_{\mu\nu}$, where

\begin{equation}
T^a_{\mu\nu}=\left[F_{\mu\rho}^aF_\nu^{a\rho}+2\frac{dV}{dA^{a2}}A^a_\mu A^a_\nu-g_{\mu\nu}\left(\frac{1}{4}F_{\rho\sigma}^{a\rho\sigma}+V(A^2) \right)\right].
\label{eq:Tmunu}
\end{equation}

\noindent Varying (\ref{LVDE}) with respect to the fields $A_\mu^a$ gives the equations of motion 

 \begin{equation}
\partial_\mu\left(\sqrt{-g}F^{a\mu\nu}\right)=2\sqrt{-g}V'A^{a\nu}, 
\label{eq:eom}
\end{equation}
\noindent  where from now on we use $V'\equiv\frac{dV}{dA^{a2}} $. 

	In an expanding universe, with FLRW metric and scale factor $a$, each one of the three vectors should be along a coordinate axis with same magnitude.  An ansatz for the $i$ components  of the vector $A^a_\mu$ compatible with homogeneity and isotropy is

\begin{equation}
A_i^a=\delta^a_i A(t),
\label{eq:ansatz}
\end{equation}

\noindent where a scalar product with an unit vector is implicit. From (\ref{eq:eom}) the component $A_0^a$ is zero and using (\ref{eq:ansatz}) into (\ref{eq:eom}) the equation of motion becomes

\begin{equation}\label{eqmotionvec}
  \ddot{A}+H\dot{A}+2V'A=0.
\end{equation} 

The pressure and energy density for the cosmic triad is obtained from (\ref{eq:Tmunu})

\begin{equation}\label{rhoA}
\rho_A=\frac{3\dot{A}^2}{2a^2}+3V,
\end{equation} 

 \begin{equation}\label{pA}
 p_A=\frac{\dot{A}^2}{2a^2}-3V+2V'\frac{A^2}{a^2}.
\end{equation} 

\noindent With this ansatz\footnote{In \cite{ArmendarizPicon:2004pm} the author used a comoving vector ansatz: $A_\mu^a=\delta_\mu^a A(t)\cdot a$. This choice leads, of course, to a different equation of motion, energy density, and pressure. However, the effect due to the scale factor that here appears in the denominator of $\rho_A$ and $p_A$, for instance, appears as a Hubble friction term ($H\dot{A}$) in the same expressions.} the potential depends now on $V(3A^2/a^2)$ and the prime is the derivative with respect to $3A^2/a^2$. We assume that the potential is given by $V=V_0 e^{-\frac{3\lambda A^2}{a^2}}$, where $V_0$ is a constant. With this form the quantity $-V'/V$ will be constant, as we will see soon. Thus, for the comoving vector $A_{ic}^a=A^a_i\cdot a$ (as used in \cite{ArmendarizPicon:2004pm}), the potential does not have an explicit dependence on the scale factor. If the cosmic triad were massless, we would have $\dot{A}\propto a^{-1}$, thus $\rho_A \propto a^{-4}$, as it should be for relativistic matter.

As we have said, we assume the  interaction between the cosmic triad with a barotropic fluid given by  $3Q \rho_m\dot{A}/a$, thus the right-hand side of Eq. (\ref{eqmotionvec}) becomes  $Q \rho_m a$. In the presence of a barotropic fluid, the Friedmann equations are

\begin{equation}\label{eq:1stFEmatterS}
  H^2=\frac{1}{3}\left(\frac{3\dot{A}^2}{2a^2}+3V+ \rho_m\right),
\end{equation}

\begin{equation}\label{eq:2ndFEmatterS}
  \dot{H}=-\frac{1}{2}\left(\frac{2\dot{A}^2}{a^2}+2V'\frac{A^2}{a^2}+(1+w_m)\rho_m\right).
\end{equation}

We now proceed to the dynamical analysis of the system.

\subsection{Autonomous system} 

The dimensionless variables are defined as

\begin{eqnarray}\label{eq:dimensionlessXYS}
 x\equiv  &\frac{\dot{A}}{\sqrt{2}Ha},  \quad y\equiv \frac{\sqrt{V(\phi)}}{H},\quad
   z\equiv \frac{A}{a},\nonumber\\ & \quad \lambda\equiv -\frac{V'}{V}, \quad \Gamma\equiv \frac{VV''}{V'^2}.
\end{eqnarray}

The dark energy density parameter is written in terms of these new variables as

\begin{equation}\label{eq:densityparameterXYS}
 \Omega_A \equiv \frac{\rho_A}{3H^2} =x^2+y^2,
 \end{equation}

\noindent so that Eq. (\ref{eq:1stFEmatterS}) can be written as 

\begin{equation}\label{eq:SomaOmegasS}
\Omega_A+\Omega_m=1,
\end{equation}

\noindent where the density parameter of the barotropic fluid is defined by $\Omega_m=\rho_m/(3H^2)$. From Eqs. (\ref{eq:densityparameterXYS}) and (\ref{eq:SomaOmegasS}) $x$ and $y$ are restricted in the phase plane by the relation

\begin{equation}\label{restrictionS}
0\leq  x^2+y^2\leq 1,
 \end{equation}
 
\noindent due to $0\leq \Omega_A\leq 1$. 

The equation of state $w_A=p_A/\rho_A$  becomes

\begin{equation}\label{eq:equationStateXYS}
 w_A =\frac{ x^2-3y^2-2\lambda z^2}{3x^2+3y^2}.
\end{equation}

\noindent Depending on the value of $\lambda$ the equation of state can be less than minus one. 

The total effective equation of state is

\begin{eqnarray}\label{eq:weffS}
 w_{eff} = \frac{p_A+p_m}{\rho_A+\rho_m}=&w_m+x^2(\frac{1}{3}-w_m)-y^2(1+w_m)\nonumber\\
&-\frac{2}{3}\lambda y^2z^2,
\end{eqnarray}

\noindent with an accelerated expansion for  $w_{eff} < -1/3$.  The dynamical system for the variables  $x$, $y$, $z$  and $\lambda$ are

\begin{eqnarray}\label{dynsystemS}\label{eq:dx1/dnS}
\frac{dx}{dN}=&-x+\sqrt{2} y^2z\lambda-\frac{3}{\sqrt{2}} Q(1-x^2-y^2)\nonumber\\
&-x\left[yz^2\lambda-x^2+y^2-\frac{1+3w_m}{2}(1-x^2-y^2)\right],\end{eqnarray}

\begin{eqnarray}\label{eq:dy/dnS}
\frac{dy}{dN}=&-3 yz\lambda(\sqrt{2}x-z)\nonumber\\&-y\left[y^2z^2\lambda-2x^2-\frac{3}{2}(1+w_m)(1-x^2-y^2)\right],
\end{eqnarray}

\begin{equation}\label{eq:dz/dnS}
\frac{dz}{dN}=2x-z,
\end{equation}

\begin{equation}\label{eq:dlambda/dnS}
\frac{d\lambda}{dN}=-6\lambda^2 z\left(\Gamma-1\right)(\sqrt{2}x-z).
\end{equation}

\subsection{Critical points}

 The fixed points of the system are obtained by setting $dx/dN=0$,  $dy/dN=0$, $dz/dN=0$ and $d\lambda/dN=0$ in Eq. (\ref{dynsystemS})--(\ref{eq:dlambda/dnS}). When $\Gamma=1$, $\lambda$ is constant the potential is $V(A^2)=V_0e^{\frac{-3\lambda A^2}{a^2}}$. \footnote{The equation for $\lambda$ is also equal zero when $z=0$ or $\lambda=0$, so that  $\lambda$ should not necessarily be constant, for the fixed point with this value of $z$. However, for the case of dynamical $\lambda$, the correspondent eigenvalue is equal to zero, indicating that the  fixed points are not hyperbolic.} Different from the scalar field case, where $V=V_0 e^{-\lambda\phi}$, the exponent of the potential also depends on the scale factor $a$. The fixed points are shown in Table \ref{criticalpointsS} with the eigenvalues of the Jacobian matrix. Notice that $y$ cannot be negative.

 \begin{table*}\centering
\begin{tabular}{llllllllll}
\hline\noalign{\smallskip}
Point  & $x$& $y$  &$z$& $w_A$ & $\Omega_A$& $w_{eff}$ & $\mu_1$ & $\mu_2$ & $\mu_3$\\
\noalign{\smallskip}\hline\noalign{\smallskip}
(a)  &$\pm 1$   &$0 $  &$\pm 2$  &$\frac{1}{3}(1-8\lambda)$&$1$ &$\frac{1}{3}$& $-1$ & $1\pm 3(\sqrt{2}Q-w_m)$&	$2+6(2-\sqrt{2})\lambda$   \\
								(b) &$\frac{\sqrt{2}Q}{w_m-1/3} $ & $0$& $2x$ &$ \frac{1}{3}(1-8\lambda)$& $\frac{2Q^2}{(w_m-1/3)^2}$ &$6Q^2$&$-1$ &$-\frac{1}{2}+9Q^2$ &$ \frac{3}{2}+Q^2(9+108(2-\sqrt{2})\lambda)$\\
		(c) &$ 0$ &$1$ &$0$ & $-1$ &$1$ &$-1$ &$-3(1+w_m)$& 	$-\frac{3}{2}(1+\sqrt{1+8\sqrt{2}\lambda})$&	$-\frac{3}{2}(1-\sqrt{1+8\sqrt{2}\lambda})$\\
   \noalign{\smallskip}\hline
\end{tabular}
\caption{\label{criticalpointsS} Critical points ($x$, $y$ and $z$) of  Eq. (\ref{dynsystemS}) for the vector-like dark energy.  The table shows the correspondent equation of state for the dark energy (\ref{eq:equationStateXYS}), the effective equation of state (\ref{eq:weffS}), the density parameter for dark energy (\ref{eq:densityparameterXYS}) and the eigenvalues of the Jacobian matrix in Eq. (\ref{Jacobian}).}
\end{table*}

The point (a) corresponds to a radiation solution, once $w_{eff}=1/3$. It can be a saddle or a stable point, depending on the value of $Q$ and $\lambda$. However, the universe is dominated by the cosmic triad, as indicated by $\Omega_A=1$, and therefore the fixed point does not describe a radiation-dominated universe, since $\Omega_m=0$. The point (b) is valid only for $w_m\neq 1/3$ and it is a saddle point,  since two eigenvalues are negative and one is positive. Since $y=0$ for this critical point, $x^2$ should be less than or equal to one (since $\Omega_A\leq 1$), so the coupling should be  $Q\leq 1/\sqrt{2}$. However, this critical point can describe a matter-dominated universe only if $Q=0$ or sufficiently small $Q\ll 1$, so that  $w_{eff} \approx 0$, as so for $\Omega_A$. The last fixed point (c) is an attractor and describes a dark-energy dominated universe ($\Omega_A=1$) that leads to an accelerated expansion of the universe, since $w_A=w_{eff}=-1$. It is a stable spiral if $\lambda<-1/(8\sqrt{2})$, otherwise it is a saddle point. The potential for this condition for $\lambda$ is $V=V_0 e^{3|\lambda|z^2}$ and it behaves as the cosmological constant at the fixed point, since $z\equiv A/a=0$ for (c). Once the coupling is constant and sufficiently small (to the fixed point (b) describe the matter-dominated universe), it has the same value, of course, for the point (c). 

\begin{figure}%
\includegraphics[width=\columnwidth]{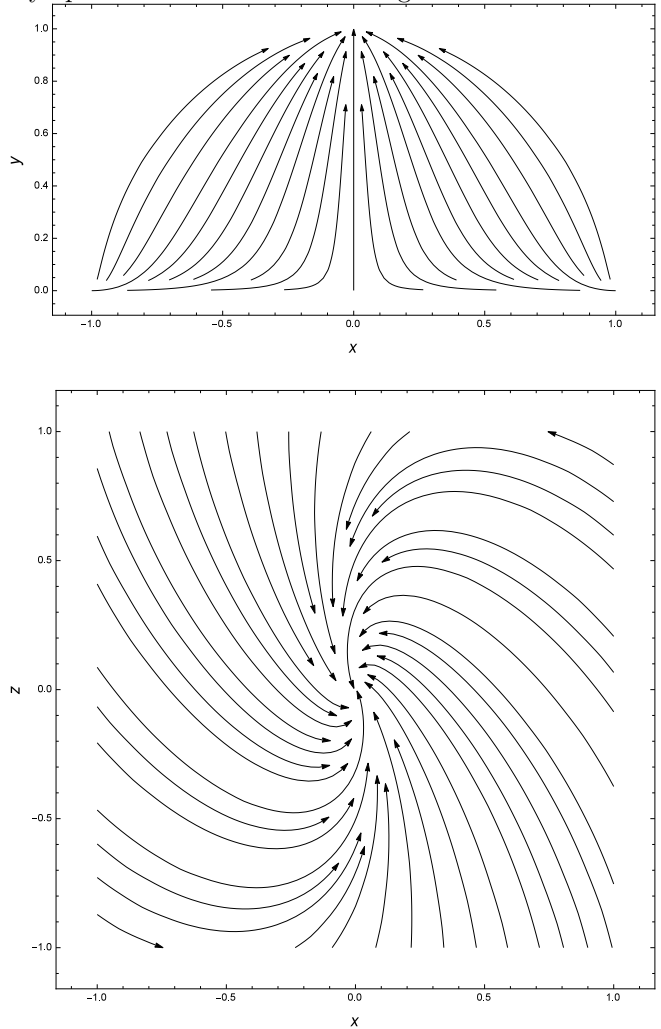}%
\caption{Phase portrait of the system, with $Q=0$ and $\lambda=-0.3$. All trajectories converge to the attractor (c) at $x=0$, $y=1$ and $z=0$, which is a stable spiral that describes the dark-energy dominated universe. The top panel shows the slice $z=0$, while the bottom panel shows the phase plane at $y=1$. }%
\label{phaseport}%
\end{figure}

\begin{figure}%
\includegraphics[width=\columnwidth]{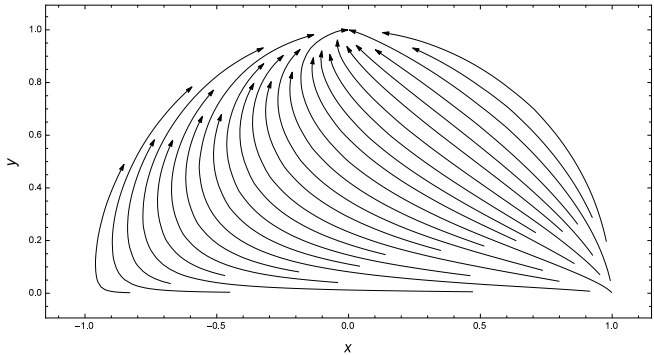}%
\caption{Phase portrait of the system, with $Q=1/\sqrt{6}$ and $\lambda=-0.3$. All trajectories converge to the attractor (c) at $x=0$, $y=1$ and $z=0$, which is a stable spiral that describes the dark-energy dominated universe. The panel shows only the slice $z=0$ because the phase plane for $y=1$ is similar to that one showed in Figure \ref{phaseport}. }%
\label{phaseport2}%
\end{figure}

We show the phase portrait of the system in Figures \ref{phaseport} ($Q=0$) and \ref{phaseport2} ($Q=1/\sqrt{6}$). The latter case is shown just to illustrate how the interaction affects the phase portrait, although we expect a very small $Q$, as discussed for the fixed point (b). We see that all trajectories converge to the attractor point.

\begin{figure}%
\includegraphics[scale=0.8]{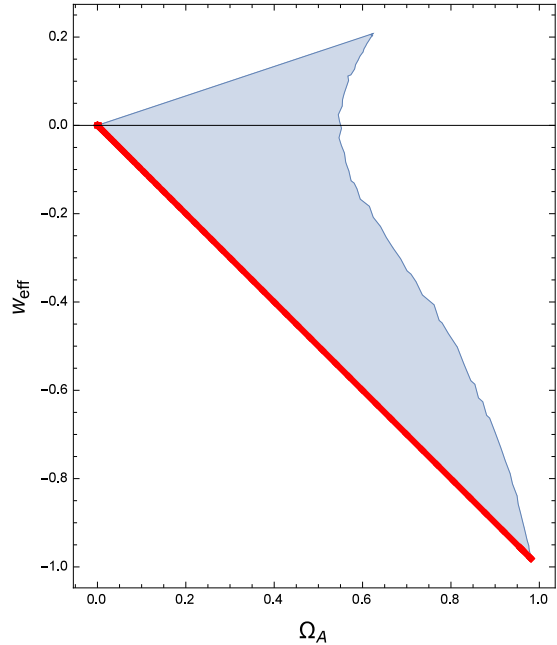}%
\caption{Effective equation of state  $w_{eff}$ (\ref{eq:weffS}) as a function of the dark energy density parameter $\Omega_A$ (\ref{eq:densityparameterXYS}). This parametric plot is independent of $Q$, once both $w_{eff}$ and $\Omega_A$  have no explicit dependence on the interaction. The blue shaded region represents the allowed values of  $w_{eff}$ and $\Omega_A$. We used $w_m=0$ since this is the only allowed value for the fixed point (b). The red line shows the transition from the fixed point (b) ($\Omega_A=0$) to the fixed point (c)  ($\Omega_A=1$). }%
\label{parametric}%
\end{figure}

In Figure \ref{parametric} we show the effective equation of state  $w_{eff}$ (\ref{eq:weffS}) as a function of the dark energy density parameter $\Omega_A$ (\ref{eq:densityparameterXYS}), where the blue shaded region represents the allowed values of  $w_{eff}$ and $\Omega_A$. The red line shows the transition from the fixed point (b) ($\Omega_A=0$) to the fixed point (c)  ($\Omega_A=1$).   

 \section{Conclusions}\label{conclu}
 
 In this paper we used the dynamical system theory to investigate if a vector-like dark energy, similar to \cite{ArmendarizPicon:2004pm}, in the presence of a barotropic fluid can lead to the three cosmological eras, namely, radiation, matter and dark energy. The analysis was generalized for the case of coupled dark energy, with a phenomenological interaction $3Q\dot{A}\rho_m/a$. There are fixed points that successfully describe the matter-dominated and the dark-energy-dominated universe. Only the radiation era was not cosmologically viable, however, if one is interested in the last two periods of the evolution of the universe, the dynamical system theory provides a good	tool to analyze asymptotic states of such cosmological models.

\begin{acknowledgments}
 This work is supported by FAPESP Grant No. 2013/10242-1. 
\end{acknowledgments}

\bibliography{trab1}

\providecommand{\noopsort}[1]{}\providecommand{\singleletter}[1]{#1}%
\begin{thebibliography}{55}%
\makeatletter
\providecommand \@ifxundefined [1]{%
 \@ifx{#1\undefined}
}%
\providecommand \@ifnum [1]{%
 \ifnum #1\expandafter \@firstoftwo
 \else \expandafter \@secondoftwo
 \fi
}%
\providecommand \@ifx [1]{%
 \ifx #1\expandafter \@firstoftwo
 \else \expandafter \@secondoftwo
 \fi
}%
\providecommand \natexlab [1]{#1}%
\providecommand \enquote  [1]{``#1''}%
\providecommand \bibnamefont  [1]{#1}%
\providecommand \bibfnamefont [1]{#1}%
\providecommand \citenamefont [1]{#1}%
\providecommand \href@noop [0]{\@secondoftwo}%
\providecommand \href [0]{\begingroup \@sanitize@url \@href}%
\providecommand \@href[1]{\@@startlink{#1}\@@href}%
\providecommand \@@href[1]{\endgroup#1\@@endlink}%
\providecommand \@sanitize@url [0]{\catcode `\\12\catcode `\$12\catcode
  `\&12\catcode `\#12\catcode `\^12\catcode `\_12\catcode `\%12\relax}%
\providecommand \@@startlink[1]{}%
\providecommand \@@endlink[0]{}%
\providecommand \url  [0]{\begingroup\@sanitize@url \@url }%
\providecommand \@url [1]{\endgroup\@href {#1}{\urlprefix }}%
\providecommand \urlprefix  [0]{URL }%
\providecommand \Eprint [0]{\href }%
\providecommand \doibase [0]{http://dx.doi.org/}%
\providecommand \selectlanguage [0]{\@gobble}%
\providecommand \bibinfo  [0]{\@secondoftwo}%
\providecommand \bibfield  [0]{\@secondoftwo}%
\providecommand \translation [1]{[#1]}%
\providecommand \BibitemOpen [0]{}%
\providecommand \bibitemStop [0]{}%
\providecommand \bibitemNoStop [0]{.\EOS\space}%
\providecommand \EOS [0]{\spacefactor3000\relax}%
\providecommand \BibitemShut  [1]{\csname bibitem#1\endcsname}%
\let\auto@bib@innerbib\@empty
\bibitem [{\citenamefont {Riess}\ \emph {et~al.}(1998)\citenamefont {Riess}
  \emph {et~al.}}]{reiss1998}%
  \BibitemOpen
  \bibfield  {author} {\bibinfo {author} {\bibfnamefont {A.~G.}\ \bibnamefont
  {Riess}} \emph {et~al.} (\bibinfo {collaboration} {Supernova Search Team}),\
  }\href {\doibase 10.1086/300499} {\bibfield  {journal} {\bibinfo  {journal}
  {Astron.J.}\ }\textbf {\bibinfo {volume} {116}},\ \bibinfo {pages} {1009}
  (\bibinfo {year} {1998})},\ \Eprint {http://arxiv.org/abs/astro-ph/9805201}
  {arXiv:astro-ph/9805201 [astro-ph]} \BibitemShut {NoStop}%
\bibitem [{\citenamefont {Perlmutter}\ \emph {et~al.}(1999)\citenamefont
  {Perlmutter} \emph {et~al.}}]{perlmutter1999}%
  \BibitemOpen
  \bibfield  {author} {\bibinfo {author} {\bibfnamefont {S.}~\bibnamefont
  {Perlmutter}} \emph {et~al.} (\bibinfo {collaboration} {Supernova Cosmology
  Project}),\ }\href {\doibase 10.1086/307221} {\bibfield  {journal} {\bibinfo
  {journal} {Astrophys.J.}\ }\textbf {\bibinfo {volume} {517}},\ \bibinfo
  {pages} {565} (\bibinfo {year} {1999})},\ \Eprint
  {http://arxiv.org/abs/astro-ph/9812133} {arXiv:astro-ph/9812133 [astro-ph]}
  \BibitemShut {NoStop}%
\bibitem [{\citenamefont {Ade}\ \emph {et~al.}(2014)\citenamefont {Ade} \emph
  {et~al.}}]{Planck2013cosmological}%
  \BibitemOpen
  \bibfield  {author} {\bibinfo {author} {\bibfnamefont {P.~A.~R.}\
  \bibnamefont {Ade}} \emph {et~al.} (\bibinfo {collaboration} {Planck}),\
  }\href {\doibase 10.1051/0004-6361/201321591} {\bibfield  {journal} {\bibinfo
   {journal} {Astron.Astrophys.}\ }\textbf {\bibinfo {volume} {571}},\ \bibinfo
  {pages} {A16} (\bibinfo {year} {2014})},\ \Eprint
  {http://arxiv.org/abs/1303.5076} {arXiv:1303.5076 [astro-ph.CO]} \BibitemShut
  {NoStop}%
\bibitem [{\citenamefont {Kachru}\ \emph {et~al.}(2003)\citenamefont {Kachru},
  \citenamefont {Kallosh}, \citenamefont {Linde},\ and\ \citenamefont
  {Trivedi}}]{kklt2003}%
  \BibitemOpen
  \bibfield  {author} {\bibinfo {author} {\bibfnamefont {S.}~\bibnamefont
  {Kachru}}, \bibinfo {author} {\bibfnamefont {R.}~\bibnamefont {Kallosh}},
  \bibinfo {author} {\bibfnamefont {A.}~\bibnamefont {Linde}}, \ and\ \bibinfo
  {author} {\bibfnamefont {S.~P.}\ \bibnamefont {Trivedi}},\ }\href@noop {}
  {\bibfield  {journal} {\bibinfo  {journal} {Phys. Rev. D}\ }\textbf {\bibinfo
  {volume} {68}},\ \bibinfo {pages} {046005} (\bibinfo {year}
  {2003})}\BibitemShut {NoStop}%
\bibitem [{\citenamefont {Copeland}\ \emph {et~al.}(2006)\citenamefont
  {Copeland}, \citenamefont {Sami},\ and\ \citenamefont
  {Tsujikawa}}]{copeland2006dynamics}%
  \BibitemOpen
  \bibfield  {author} {\bibinfo {author} {\bibfnamefont {E.~J.}\ \bibnamefont
  {Copeland}}, \bibinfo {author} {\bibfnamefont {M.}~\bibnamefont {Sami}}, \
  and\ \bibinfo {author} {\bibfnamefont {S.}~\bibnamefont {Tsujikawa}},\ }\href
  {\doibase 10.1142/S021827180600942X} {\bibfield  {journal} {\bibinfo
  {journal} {Int. J. Mod. Phys.}\ }\textbf {\bibinfo {volume} {D15}},\ \bibinfo
  {pages} {1753} (\bibinfo {year} {2006})},\ \Eprint
  {http://arxiv.org/abs/hep-th/0603057} {arXiv:hep-th/0603057 [hep-th]}
  \BibitemShut {NoStop}%
\bibitem [{\citenamefont {Dvali}\ \emph {et~al.}(2000)\citenamefont {Dvali},
  \citenamefont {Gabadadze},\ and\ \citenamefont {Porrati}}]{dvali2000}%
  \BibitemOpen
  \bibfield  {author} {\bibinfo {author} {\bibfnamefont {G.}~\bibnamefont
  {Dvali}}, \bibinfo {author} {\bibfnamefont {G.}~\bibnamefont {Gabadadze}}, \
  and\ \bibinfo {author} {\bibfnamefont {M.}~\bibnamefont {Porrati}},\
  }\href@noop {} {\bibfield  {journal} {\bibinfo  {journal} {Phys. Lett. B}\
  }\textbf {\bibinfo {volume} {485}},\ \bibinfo {pages} {208} (\bibinfo {year}
  {2000})}\BibitemShut {NoStop}%
\bibitem [{\citenamefont {Yin}\ \emph {et~al.}(2007)\citenamefont {Yin},
  \citenamefont {Wang}, \citenamefont {Abdalla},\ and\ \citenamefont
  {Lin}}]{yin2005}%
  \BibitemOpen
  \bibfield  {author} {\bibinfo {author} {\bibfnamefont {S.}~\bibnamefont
  {Yin}}, \bibinfo {author} {\bibfnamefont {B.}~\bibnamefont {Wang}}, \bibinfo
  {author} {\bibfnamefont {E.}~\bibnamefont {Abdalla}}, \ and\ \bibinfo
  {author} {\bibfnamefont {C.}~\bibnamefont {Lin}},\ }\href@noop {} {\bibfield
  {journal} {\bibinfo  {journal} {Phys.Rev. D}\ }\textbf {\bibinfo {volume}
  {76}},\ \bibinfo {pages} {124026} (\bibinfo {year} {2007})}\BibitemShut
  {NoStop}%
\bibitem [{\citenamefont {Peebles}\ and\ \citenamefont
  {Ratra}(1988)}]{peebles1988}%
  \BibitemOpen
  \bibfield  {author} {\bibinfo {author} {\bibfnamefont {P.~J.~E.}\
  \bibnamefont {Peebles}}\ and\ \bibinfo {author} {\bibfnamefont
  {B.}~\bibnamefont {Ratra}},\ }\href {\doibase 10.1086/185100} {\bibfield
  {journal} {\bibinfo  {journal} {Astrophys.J.}\ }\textbf {\bibinfo {volume}
  {325}},\ \bibinfo {pages} {L17} (\bibinfo {year} {1988})}\BibitemShut
  {NoStop}%
\bibitem [{\citenamefont {Ratra}\ and\ \citenamefont
  {Peebles}(1988)}]{ratra1988}%
  \BibitemOpen
  \bibfield  {author} {\bibinfo {author} {\bibfnamefont {B.}~\bibnamefont
  {Ratra}}\ and\ \bibinfo {author} {\bibfnamefont {P.~J.~E.}\ \bibnamefont
  {Peebles}},\ }\href {\doibase 10.1103/PhysRevD.37.3406} {\bibfield  {journal}
  {\bibinfo  {journal} {Phys.Rev.}\ }\textbf {\bibinfo {volume} {D37}},\
  \bibinfo {pages} {3406} (\bibinfo {year} {1988})}\BibitemShut {NoStop}%
\bibitem [{\citenamefont {Frieman}\ \emph {et~al.}(1992)\citenamefont
  {Frieman}, \citenamefont {Hill},\ and\ \citenamefont
  {Watkins}}]{Frieman1992}%
  \BibitemOpen
  \bibfield  {author} {\bibinfo {author} {\bibfnamefont {J.~A.}\ \bibnamefont
  {Frieman}}, \bibinfo {author} {\bibfnamefont {C.~T.}\ \bibnamefont {Hill}}, \
  and\ \bibinfo {author} {\bibfnamefont {R.}~\bibnamefont {Watkins}},\ }\href
  {\doibase 10.1103/PhysRevD.46.1226} {\bibfield  {journal} {\bibinfo
  {journal} {Phys.Rev.}\ }\textbf {\bibinfo {volume} {D46}},\ \bibinfo {pages}
  {1226} (\bibinfo {year} {1992})}\BibitemShut {NoStop}%
\bibitem [{\citenamefont {Frieman}\ \emph {et~al.}(1995)\citenamefont
  {Frieman}, \citenamefont {Hill}, \citenamefont {Stebbins},\ and\
  \citenamefont {Waga}}]{Frieman1995}%
  \BibitemOpen
  \bibfield  {author} {\bibinfo {author} {\bibfnamefont {J.~A.}\ \bibnamefont
  {Frieman}}, \bibinfo {author} {\bibfnamefont {C.~T.}\ \bibnamefont {Hill}},
  \bibinfo {author} {\bibfnamefont {A.}~\bibnamefont {Stebbins}}, \ and\
  \bibinfo {author} {\bibfnamefont {I.}~\bibnamefont {Waga}},\ }\href {\doibase
  10.1103/PhysRevLett.80.1582} {\bibfield  {journal} {\bibinfo  {journal}
  {Phys. Rev. Lett.}\ }\textbf {\bibinfo {volume} {75}},\ \bibinfo {pages}
  {2077} (\bibinfo {year} {1995})}\BibitemShut {NoStop}%
\bibitem [{\citenamefont {Caldwell}\ \emph {et~al.}(1998)\citenamefont
  {Caldwell}, \citenamefont {Dave},\ and\ \citenamefont
  {Steinhardt}}]{Caldwell:1997ii}%
  \BibitemOpen
  \bibfield  {author} {\bibinfo {author} {\bibfnamefont {R.~R.}\ \bibnamefont
  {Caldwell}}, \bibinfo {author} {\bibfnamefont {R.}~\bibnamefont {Dave}}, \
  and\ \bibinfo {author} {\bibfnamefont {P.~J.}\ \bibnamefont {Steinhardt}},\
  }\href {\doibase 10.1103/PhysRevLett.80.1582} {\bibfield  {journal} {\bibinfo
   {journal} {Phys. Rev. Lett.}\ }\textbf {\bibinfo {volume} {80}},\ \bibinfo
  {pages} {1582} (\bibinfo {year} {1998})}\BibitemShut {NoStop}%
\bibitem [{\citenamefont {Padmanabhan}(2002)}]{Padmanabhan:2002cp}%
  \BibitemOpen
  \bibfield  {author} {\bibinfo {author} {\bibfnamefont {T.}~\bibnamefont
  {Padmanabhan}},\ }\href {\doibase 10.1103/PhysRevD.66.021301} {\bibfield
  {journal} {\bibinfo  {journal} {Phys.Rev.}\ }\textbf {\bibinfo {volume}
  {D66}},\ \bibinfo {pages} {021301} (\bibinfo {year} {2002})},\ \Eprint
  {http://arxiv.org/abs/hep-th/0204150} {arXiv:hep-th/0204150 [hep-th]}
  \BibitemShut {NoStop}%
\bibitem [{\citenamefont {Bagla}\ \emph {et~al.}(2003)\citenamefont {Bagla},
  \citenamefont {Jassal},\ and\ \citenamefont {Padmanabhan}}]{Bagla:2002yn}%
  \BibitemOpen
  \bibfield  {author} {\bibinfo {author} {\bibfnamefont {J.~S.}\ \bibnamefont
  {Bagla}}, \bibinfo {author} {\bibfnamefont {H.~K.}\ \bibnamefont {Jassal}}, \
  and\ \bibinfo {author} {\bibfnamefont {T.}~\bibnamefont {Padmanabhan}},\
  }\href {\doibase 10.1103/PhysRevD.67.063504} {\bibfield  {journal} {\bibinfo
  {journal} {Phys.Rev.}\ }\textbf {\bibinfo {volume} {D67}},\ \bibinfo {pages}
  {063504} (\bibinfo {year} {2003})},\ \Eprint
  {http://arxiv.org/abs/astro-ph/0212198} {arXiv:astro-ph/0212198 [astro-ph]}
  \BibitemShut {NoStop}%
\bibitem [{\citenamefont {Armendariz-Picon}\ \emph {et~al.}(2000)\citenamefont
  {Armendariz-Picon}, \citenamefont {Mukhanov},\ and\ \citenamefont
  {Steinhardt}}]{ArmendarizPicon:2000dh}%
  \BibitemOpen
  \bibfield  {author} {\bibinfo {author} {\bibfnamefont {C.}~\bibnamefont
  {Armendariz-Picon}}, \bibinfo {author} {\bibfnamefont {V.~F.}\ \bibnamefont
  {Mukhanov}}, \ and\ \bibinfo {author} {\bibfnamefont {P.~J.}\ \bibnamefont
  {Steinhardt}},\ }\href {\doibase 10.1103/PhysRevLett.85.4438} {\bibfield
  {journal} {\bibinfo  {journal} {Phys. Rev. Lett.}\ }\textbf {\bibinfo
  {volume} {85}},\ \bibinfo {pages} {4438} (\bibinfo {year} {2000})},\ \Eprint
  {http://arxiv.org/abs/astro-ph/0004134} {arXiv:astro-ph/0004134 [astro-ph]}
  \BibitemShut {NoStop}%
\bibitem [{\citenamefont {Brax}\ and\ \citenamefont {Martin}(1999)}]{Brax1999}%
  \BibitemOpen
  \bibfield  {author} {\bibinfo {author} {\bibfnamefont {P.}~\bibnamefont
  {Brax}}\ and\ \bibinfo {author} {\bibfnamefont {J.}~\bibnamefont {Martin}},\
  }\href {\doibase 10.1016/S0370-2693(99)01209-5} {\bibfield  {journal}
  {\bibinfo  {journal} {Phys. Lett.}\ }\textbf {\bibinfo {volume} {B468}},\
  \bibinfo {pages} {40} (\bibinfo {year} {1999})},\ \Eprint
  {http://arxiv.org/abs/astro-ph/9905040} {arXiv:astro-ph/9905040 [astro-ph]}
  \BibitemShut {NoStop}%
\bibitem [{\citenamefont {Copeland}\ \emph {et~al.}(2000)\citenamefont
  {Copeland}, \citenamefont {Nunes},\ and\ \citenamefont
  {Rosati}}]{Copeland2000}%
  \BibitemOpen
  \bibfield  {author} {\bibinfo {author} {\bibfnamefont {E.~J.}\ \bibnamefont
  {Copeland}}, \bibinfo {author} {\bibfnamefont {N.~J.}\ \bibnamefont {Nunes}},
  \ and\ \bibinfo {author} {\bibfnamefont {F.}~\bibnamefont {Rosati}},\ }\href
  {\doibase 10.1103/PhysRevD.62.123503} {\bibfield  {journal} {\bibinfo
  {journal} {Phys. Rev.}\ }\textbf {\bibinfo {volume} {D62}},\ \bibinfo {pages}
  {123503} (\bibinfo {year} {2000})},\ \Eprint
  {http://arxiv.org/abs/hep-ph/0005222} {arXiv:hep-ph/0005222 [hep-ph]}
  \BibitemShut {NoStop}%
\bibitem [{\citenamefont {Landim}(2016{\natexlab{a}})}]{Landim:2015upa}%
  \BibitemOpen
  \bibfield  {author} {\bibinfo {author} {\bibfnamefont {R.~C.~G.}\
  \bibnamefont {Landim}},\ }\href {\doibase 10.1140/epjc/s10052-016-4287-2}
  {\bibfield  {journal} {\bibinfo  {journal} {Eur. Phys. J.}\ }\textbf
  {\bibinfo {volume} {C76}},\ \bibinfo {pages} {430} (\bibinfo {year}
  {2016}{\natexlab{a}})},\ \Eprint {http://arxiv.org/abs/1509.04980}
  {arXiv:1509.04980 [hep-th]} \BibitemShut {NoStop}%
\bibitem [{\citenamefont {Landim}(2016{\natexlab{b}})}]{Landim:2015hqa}%
  \BibitemOpen
  \bibfield  {author} {\bibinfo {author} {\bibfnamefont {R.~C.~G.}\
  \bibnamefont {Landim}},\ }\href {\doibase 10.1142/S0218271816500504}
  {\bibfield  {journal} {\bibinfo  {journal} {Int. J. Mod. Phys.}\ }\textbf
  {\bibinfo {volume} {D25}},\ \bibinfo {pages} {1650050} (\bibinfo {year}
  {2016}{\natexlab{b}})},\ \Eprint {http://arxiv.org/abs/1508.07248}
  {arXiv:1508.07248 [hep-th]} \BibitemShut {NoStop}%
\bibitem [{\citenamefont {Armendariz-Picon}(2004)}]{ArmendarizPicon:2004pm}%
  \BibitemOpen
  \bibfield  {author} {\bibinfo {author} {\bibfnamefont {C.}~\bibnamefont
  {Armendariz-Picon}},\ }\href {\doibase 10.1088/1475-7516/2004/07/007}
  {\bibfield  {journal} {\bibinfo  {journal} {JCAP}\ }\textbf {\bibinfo
  {volume} {0407}},\ \bibinfo {pages} {007} (\bibinfo {year} {2004})},\ \Eprint
  {http://arxiv.org/abs/astro-ph/0405267} {arXiv:astro-ph/0405267 [astro-ph]}
  \BibitemShut {NoStop}%
\bibitem [{\citenamefont {Koivisto}\ and\ \citenamefont
  {Mota}(2008)}]{Koivisto:2008xf}%
  \BibitemOpen
  \bibfield  {author} {\bibinfo {author} {\bibfnamefont {T.}~\bibnamefont
  {Koivisto}}\ and\ \bibinfo {author} {\bibfnamefont {D.~F.}\ \bibnamefont
  {Mota}},\ }\href {\doibase 10.1088/1475-7516/2008/08/021} {\bibfield
  {journal} {\bibinfo  {journal} {JCAP}\ }\textbf {\bibinfo {volume} {0808}},\
  \bibinfo {pages} {021} (\bibinfo {year} {2008})},\ \Eprint
  {http://arxiv.org/abs/0805.4229} {arXiv:0805.4229 [astro-ph]} \BibitemShut
  {NoStop}%
\bibitem [{\citenamefont {Bamba}\ and\ \citenamefont
  {Odintsov}(2008)}]{Bamba:2008ja}%
  \BibitemOpen
  \bibfield  {author} {\bibinfo {author} {\bibfnamefont {K.}~\bibnamefont
  {Bamba}}\ and\ \bibinfo {author} {\bibfnamefont {S.~D.}\ \bibnamefont
  {Odintsov}},\ }\href {\doibase 10.1088/1475-7516/2008/04/024} {\bibfield
  {journal} {\bibinfo  {journal} {JCAP}\ }\textbf {\bibinfo {volume} {0804}},\
  \bibinfo {pages} {024} (\bibinfo {year} {2008})},\ \Eprint
  {http://arxiv.org/abs/0801.0954} {arXiv:0801.0954 [astro-ph]} \BibitemShut
  {NoStop}%
\bibitem [{\citenamefont {Emelyanov}\ and\ \citenamefont
  {Klinkhamer}(2012{\natexlab{a}})}]{Emelyanov:2011ze}%
  \BibitemOpen
  \bibfield  {author} {\bibinfo {author} {\bibfnamefont {V.}~\bibnamefont
  {Emelyanov}}\ and\ \bibinfo {author} {\bibfnamefont {F.~R.}\ \bibnamefont
  {Klinkhamer}},\ }\href {\doibase 10.1103/PhysRevD.85.103508} {\bibfield
  {journal} {\bibinfo  {journal} {Phys. Rev.}\ }\textbf {\bibinfo {volume}
  {D85}},\ \bibinfo {pages} {103508} (\bibinfo {year} {2012}{\natexlab{a}})},\
  \Eprint {http://arxiv.org/abs/1109.4915} {arXiv:1109.4915 [hep-th]}
  \BibitemShut {NoStop}%
\bibitem [{\citenamefont {Emelyanov}\ and\ \citenamefont
  {Klinkhamer}(2012{\natexlab{b}})}]{Emelyanov:2011wn}%
  \BibitemOpen
  \bibfield  {author} {\bibinfo {author} {\bibfnamefont {V.}~\bibnamefont
  {Emelyanov}}\ and\ \bibinfo {author} {\bibfnamefont {F.~R.}\ \bibnamefont
  {Klinkhamer}},\ }\href {\doibase 10.1103/PhysRevD.85.063522} {\bibfield
  {journal} {\bibinfo  {journal} {Phys. Rev.}\ }\textbf {\bibinfo {volume}
  {D85}},\ \bibinfo {pages} {063522} (\bibinfo {year} {2012}{\natexlab{b}})},\
  \Eprint {http://arxiv.org/abs/1107.0961} {arXiv:1107.0961 [hep-th]}
  \BibitemShut {NoStop}%
\bibitem [{\citenamefont {Emelyanov}\ and\ \citenamefont
  {Klinkhamer}(2012{\natexlab{c}})}]{Emelyanov:2011kn}%
  \BibitemOpen
  \bibfield  {author} {\bibinfo {author} {\bibfnamefont {V.}~\bibnamefont
  {Emelyanov}}\ and\ \bibinfo {author} {\bibfnamefont {F.~R.}\ \bibnamefont
  {Klinkhamer}},\ }\href {\doibase 10.1142/S0218271812500253} {\bibfield
  {journal} {\bibinfo  {journal} {Int. J. Mod. Phys.}\ }\textbf {\bibinfo
  {volume} {D21}},\ \bibinfo {pages} {1250025} (\bibinfo {year}
  {2012}{\natexlab{c}})},\ \Eprint {http://arxiv.org/abs/1108.1995}
  {arXiv:1108.1995 [gr-qc]} \BibitemShut {NoStop}%
\bibitem [{\citenamefont {Kouwn}\ \emph {et~al.}(2016)\citenamefont {Kouwn},
  \citenamefont {Oh},\ and\ \citenamefont {Park}}]{Kouwn:2015cdw}%
  \BibitemOpen
  \bibfield  {author} {\bibinfo {author} {\bibfnamefont {S.}~\bibnamefont
  {Kouwn}}, \bibinfo {author} {\bibfnamefont {P.}~\bibnamefont {Oh}}, \ and\
  \bibinfo {author} {\bibfnamefont {C.-G.}\ \bibnamefont {Park}},\ }\href
  {\doibase 10.1103/PhysRevD.93.083012} {\bibfield  {journal} {\bibinfo
  {journal} {Phys. Rev.}\ }\textbf {\bibinfo {volume} {D93}},\ \bibinfo {pages}
  {083012} (\bibinfo {year} {2016})},\ \Eprint
  {http://arxiv.org/abs/1512.00541} {arXiv:1512.00541 [astro-ph.CO]}
  \BibitemShut {NoStop}%
\bibitem [{\citenamefont {Wetterich}(1995)}]{Wetterich:1994bg}%
  \BibitemOpen
  \bibfield  {author} {\bibinfo {author} {\bibfnamefont {C.}~\bibnamefont
  {Wetterich}},\ }\href@noop {} {\bibfield  {journal} {\bibinfo  {journal}
  {Astron.Astrophys.}\ }\textbf {\bibinfo {volume} {301}},\ \bibinfo {pages}
  {321} (\bibinfo {year} {1995})},\ \Eprint
  {http://arxiv.org/abs/hep-th/9408025} {arXiv:hep-th/9408025 [hep-th]}
  \BibitemShut {NoStop}%
\bibitem [{\citenamefont {Amendola}(2000)}]{Amendola:1999er}%
  \BibitemOpen
  \bibfield  {author} {\bibinfo {author} {\bibfnamefont {L.}~\bibnamefont
  {Amendola}},\ }\href {\doibase 10.1103/PhysRevD.62.043511} {\bibfield
  {journal} {\bibinfo  {journal} {Phys.Rev.}\ }\textbf {\bibinfo {volume}
  {D62}},\ \bibinfo {pages} {043511} (\bibinfo {year} {2000})},\ \Eprint
  {http://arxiv.org/abs/astro-ph/9908023} {arXiv:astro-ph/9908023 [astro-ph]}
  \BibitemShut {NoStop}%
\bibitem [{\citenamefont {Farrar}\ and\ \citenamefont
  {Peebles}(2004)}]{Farrar:2003uw}%
  \BibitemOpen
  \bibfield  {author} {\bibinfo {author} {\bibfnamefont {G.~R.}\ \bibnamefont
  {Farrar}}\ and\ \bibinfo {author} {\bibfnamefont {P.~J.~E.}\ \bibnamefont
  {Peebles}},\ }\href {\doibase 10.1086/381728} {\bibfield  {journal} {\bibinfo
   {journal} {Astrophys. J.}\ }\textbf {\bibinfo {volume} {604}},\ \bibinfo
  {pages} {1} (\bibinfo {year} {2004})},\ \Eprint
  {http://arxiv.org/abs/astro-ph/0307316} {arXiv:astro-ph/0307316 [astro-ph]}
  \BibitemShut {NoStop}%
\bibitem [{\citenamefont {Guo}\ and\ \citenamefont {Zhang}(2005)}]{Guo:2004vg}%
  \BibitemOpen
  \bibfield  {author} {\bibinfo {author} {\bibfnamefont {Z.-K.}\ \bibnamefont
  {Guo}}\ and\ \bibinfo {author} {\bibfnamefont {Y.-Z.}\ \bibnamefont
  {Zhang}},\ }\href {\doibase 10.1103/PhysRevD.71.023501} {\bibfield  {journal}
  {\bibinfo  {journal} {Phys. Rev. D.}\ }\textbf {\bibinfo {volume} {71}},\
  \bibinfo {pages} {023501} (\bibinfo {year} {2005})},\ \Eprint
  {http://arxiv.org/abs/astro-ph/0411524} {arXiv:astro-ph/0411524 [astro-ph]}
  \BibitemShut {NoStop}%
\bibitem [{\citenamefont {Cai}\ and\ \citenamefont {Wang}(2005)}]{Cai:2004dk}%
  \BibitemOpen
  \bibfield  {author} {\bibinfo {author} {\bibfnamefont {R.-G.}\ \bibnamefont
  {Cai}}\ and\ \bibinfo {author} {\bibfnamefont {A.}~\bibnamefont {Wang}},\
  }\href {\doibase 10.1088/1475-7516/2005/03/002} {\bibfield  {journal}
  {\bibinfo  {journal} {JCAP}\ }\textbf {\bibinfo {volume} {0503}},\ \bibinfo
  {pages} {002} (\bibinfo {year} {2005})},\ \Eprint
  {http://arxiv.org/abs/hep-th/0411025} {arXiv:hep-th/0411025 [hep-th]}
  \BibitemShut {NoStop}%
\bibitem [{\citenamefont {Guo}\ \emph {et~al.}(2005)\citenamefont {Guo},
  \citenamefont {Cai},\ and\ \citenamefont {Zhang}}]{Guo:2004xx}%
  \BibitemOpen
  \bibfield  {author} {\bibinfo {author} {\bibfnamefont {Z.-K.}\ \bibnamefont
  {Guo}}, \bibinfo {author} {\bibfnamefont {R.-G.}\ \bibnamefont {Cai}}, \ and\
  \bibinfo {author} {\bibfnamefont {Y.-Z.}\ \bibnamefont {Zhang}},\ }\href
  {\doibase 10.1088/1475-7516/2005/05/002} {\bibfield  {journal} {\bibinfo
  {journal} {JCAP}\ }\textbf {\bibinfo {volume} {0505}},\ \bibinfo {pages}
  {002} (\bibinfo {year} {2005})},\ \Eprint
  {http://arxiv.org/abs/astro-ph/0412624} {arXiv:astro-ph/0412624 [astro-ph]}
  \BibitemShut {NoStop}%
\bibitem [{\citenamefont {Bi}\ \emph {et~al.}(2005)\citenamefont {Bi},
  \citenamefont {Feng}, \citenamefont {Li},\ and\ \citenamefont
  {Zhang}}]{Bi:2004ns}%
  \BibitemOpen
  \bibfield  {author} {\bibinfo {author} {\bibfnamefont {X.-J.}\ \bibnamefont
  {Bi}}, \bibinfo {author} {\bibfnamefont {B.}~\bibnamefont {Feng}}, \bibinfo
  {author} {\bibfnamefont {H.}~\bibnamefont {Li}}, \ and\ \bibinfo {author}
  {\bibfnamefont {X.}~\bibnamefont {Zhang}},\ }\href {\doibase
  10.1103/PhysRevD.72.123523} {\bibfield  {journal} {\bibinfo  {journal} {Phys.
  Rev. D.}\ }\textbf {\bibinfo {volume} {72}},\ \bibinfo {pages} {123523}
  (\bibinfo {year} {2005})},\ \Eprint {http://arxiv.org/abs/hep-ph/0412002}
  {arXiv:hep-ph/0412002 [hep-ph]} \BibitemShut {NoStop}%
\bibitem [{\citenamefont {Gumjudpai}\ \emph {et~al.}(2005)\citenamefont
  {Gumjudpai}, \citenamefont {Naskar}, \citenamefont {Sami},\ and\
  \citenamefont {Tsujikawa}}]{Gumjudpai:2005ry}%
  \BibitemOpen
  \bibfield  {author} {\bibinfo {author} {\bibfnamefont {B.}~\bibnamefont
  {Gumjudpai}}, \bibinfo {author} {\bibfnamefont {T.}~\bibnamefont {Naskar}},
  \bibinfo {author} {\bibfnamefont {M.}~\bibnamefont {Sami}}, \ and\ \bibinfo
  {author} {\bibfnamefont {S.}~\bibnamefont {Tsujikawa}},\ }\href {\doibase
  10.1088/1475-7516/2005/06/007} {\bibfield  {journal} {\bibinfo  {journal}
  {JCAP}\ }\textbf {\bibinfo {volume} {0506}},\ \bibinfo {pages} {007}
  (\bibinfo {year} {2005})},\ \Eprint {http://arxiv.org/abs/hep-th/0502191}
  {arXiv:hep-th/0502191 [hep-th]} \BibitemShut {NoStop}%
\bibitem [{\citenamefont {Wang}\ \emph {et~al.}(2005)\citenamefont {Wang},
  \citenamefont {Gong},\ and\ \citenamefont {Abdalla}}]{Wang:2005jx}%
  \BibitemOpen
  \bibfield  {author} {\bibinfo {author} {\bibfnamefont {B.}~\bibnamefont
  {Wang}}, \bibinfo {author} {\bibfnamefont {Y.-G.}\ \bibnamefont {Gong}}, \
  and\ \bibinfo {author} {\bibfnamefont {E.}~\bibnamefont {Abdalla}},\ }\href
  {\doibase 10.1016/j.physletb.2005.08.008} {\bibfield  {journal} {\bibinfo
  {journal} {Phys. Lett.}\ }\textbf {\bibinfo {volume} {B624}},\ \bibinfo
  {pages} {141} (\bibinfo {year} {2005})},\ \Eprint
  {http://arxiv.org/abs/hep-th/0506069} {arXiv:hep-th/0506069 [hep-th]}
  \BibitemShut {NoStop}%
\bibitem [{\citenamefont {Wang}\ \emph
  {et~al.}(2006{\natexlab{a}})\citenamefont {Wang}, \citenamefont {Gong},\ and\
  \citenamefont {Abdalla}}]{Wang:2005pk}%
  \BibitemOpen
  \bibfield  {author} {\bibinfo {author} {\bibfnamefont {B.}~\bibnamefont
  {Wang}}, \bibinfo {author} {\bibfnamefont {Y.}~\bibnamefont {Gong}}, \ and\
  \bibinfo {author} {\bibfnamefont {E.}~\bibnamefont {Abdalla}},\ }\href
  {\doibase 10.1103/PhysRevD.74.083520} {\bibfield  {journal} {\bibinfo
  {journal} {Phys. Rev.}\ }\textbf {\bibinfo {volume} {D74}},\ \bibinfo {pages}
  {083520} (\bibinfo {year} {2006}{\natexlab{a}})},\ \Eprint
  {http://arxiv.org/abs/gr-qc/0511051} {arXiv:gr-qc/0511051 [gr-qc]}
  \BibitemShut {NoStop}%
\bibitem [{\citenamefont {Wang}\ \emph
  {et~al.}(2006{\natexlab{b}})\citenamefont {Wang}, \citenamefont {Lin},\ and\
  \citenamefont {Abdalla}}]{Wang:2005ph}%
  \BibitemOpen
  \bibfield  {author} {\bibinfo {author} {\bibfnamefont {B.}~\bibnamefont
  {Wang}}, \bibinfo {author} {\bibfnamefont {C.-Y.}\ \bibnamefont {Lin}}, \
  and\ \bibinfo {author} {\bibfnamefont {E.}~\bibnamefont {Abdalla}},\ }\href
  {\doibase 10.1016/j.physletb.2006.04.009} {\bibfield  {journal} {\bibinfo
  {journal} {Phys. Lett.}\ }\textbf {\bibinfo {volume} {B637}},\ \bibinfo
  {pages} {357} (\bibinfo {year} {2006}{\natexlab{b}})},\ \Eprint
  {http://arxiv.org/abs/hep-th/0509107} {arXiv:hep-th/0509107 [hep-th]}
  \BibitemShut {NoStop}%
\bibitem [{\citenamefont {Wang}\ \emph {et~al.}(2008)\citenamefont {Wang},
  \citenamefont {Lin}, \citenamefont {Pavon},\ and\ \citenamefont
  {Abdalla}}]{Wang:2007ak}%
  \BibitemOpen
  \bibfield  {author} {\bibinfo {author} {\bibfnamefont {B.}~\bibnamefont
  {Wang}}, \bibinfo {author} {\bibfnamefont {C.-Y.}\ \bibnamefont {Lin}},
  \bibinfo {author} {\bibfnamefont {D.}~\bibnamefont {Pavon}}, \ and\ \bibinfo
  {author} {\bibfnamefont {E.}~\bibnamefont {Abdalla}},\ }\href {\doibase
  10.1016/j.physletb.2008.01.074} {\bibfield  {journal} {\bibinfo  {journal}
  {Phys. Lett.}\ }\textbf {\bibinfo {volume} {B662}},\ \bibinfo {pages} {1}
  (\bibinfo {year} {2008})},\ \Eprint {http://arxiv.org/abs/0711.2214}
  {arXiv:0711.2214 [hep-th]} \BibitemShut {NoStop}%
\bibitem [{\citenamefont {Micheletti}\ \emph {et~al.}(2009)\citenamefont
  {Micheletti}, \citenamefont {Abdalla},\ and\ \citenamefont
  {Wang}}]{micheletti2009}%
  \BibitemOpen
  \bibfield  {author} {\bibinfo {author} {\bibfnamefont {S.}~\bibnamefont
  {Micheletti}}, \bibinfo {author} {\bibfnamefont {E.}~\bibnamefont {Abdalla}},
  \ and\ \bibinfo {author} {\bibfnamefont {B.}~\bibnamefont {Wang}},\ }\href
  {\doibase 10.1103/PhysRevD.79.123506} {\bibfield  {journal} {\bibinfo
  {journal} {Phys.Rev.}\ }\textbf {\bibinfo {volume} {D79}},\ \bibinfo {pages}
  {123506} (\bibinfo {year} {2009})},\ \Eprint {http://arxiv.org/abs/0902.0318}
  {arXiv:0902.0318 [gr-qc]} \BibitemShut {NoStop}%
\bibitem [{\citenamefont {Costa}\ \emph {et~al.}(2015)\citenamefont {Costa},
  \citenamefont {Olivari},\ and\ \citenamefont {Abdalla}}]{Costa:2014pba}%
  \BibitemOpen
  \bibfield  {author} {\bibinfo {author} {\bibfnamefont {A.~A.}\ \bibnamefont
  {Costa}}, \bibinfo {author} {\bibfnamefont {L.~C.}\ \bibnamefont {Olivari}},
  \ and\ \bibinfo {author} {\bibfnamefont {E.}~\bibnamefont {Abdalla}},\ }\href
  {\doibase 10.1103/PhysRevD.92.103501} {\bibfield  {journal} {\bibinfo
  {journal} {Phys. Rev.}\ }\textbf {\bibinfo {volume} {D92}},\ \bibinfo {pages}
  {103501} (\bibinfo {year} {2015})},\ \Eprint {http://arxiv.org/abs/1411.3660}
  {arXiv:1411.3660 [astro-ph.CO]} \BibitemShut {NoStop}%
\bibitem [{\citenamefont {Wang}\ \emph {et~al.}(2016)\citenamefont {Wang},
  \citenamefont {Abdalla}, \citenamefont {Atrio-Barandela},\ and\ \citenamefont
  {Pavon}}]{Wang:2016lxa}%
  \BibitemOpen
  \bibfield  {author} {\bibinfo {author} {\bibfnamefont {B.}~\bibnamefont
  {Wang}}, \bibinfo {author} {\bibfnamefont {E.}~\bibnamefont {Abdalla}},
  \bibinfo {author} {\bibfnamefont {F.}~\bibnamefont {Atrio-Barandela}}, \ and\
  \bibinfo {author} {\bibfnamefont {D.}~\bibnamefont {Pavon}},\ }\href
  {\doibase 10.1088/0034-4885/79/9/096901} {\bibfield  {journal} {\bibinfo
  {journal} {Rept. Prog. Phys.}\ }\textbf {\bibinfo {volume} {79}},\ \bibinfo
  {pages} {096901} (\bibinfo {year} {2016})},\ \Eprint
  {http://arxiv.org/abs/1603.08299} {arXiv:1603.08299 [astro-ph.CO]}
  \BibitemShut {NoStop}%
\bibitem [{\citenamefont {Zimdahl}\ and\ \citenamefont
  {Pavon}(2001)}]{Zimdahl:2001ar}%
  \BibitemOpen
  \bibfield  {author} {\bibinfo {author} {\bibfnamefont {W.}~\bibnamefont
  {Zimdahl}}\ and\ \bibinfo {author} {\bibfnamefont {D.}~\bibnamefont
  {Pavon}},\ }\href {\doibase 10.1016/S0370-2693(01)01174-1} {\bibfield
  {journal} {\bibinfo  {journal} {Phys.Lett.}\ }\textbf {\bibinfo {volume}
  {B521}},\ \bibinfo {pages} {133} (\bibinfo {year} {2001})},\ \Eprint
  {http://arxiv.org/abs/astro-ph/0105479} {arXiv:astro-ph/0105479 [astro-ph]}
  \BibitemShut {NoStop}%
\bibitem [{\citenamefont {Chimento}\ \emph {et~al.}(2003)\citenamefont
  {Chimento}, \citenamefont {Jakubi}, \citenamefont {Pavon},\ and\
  \citenamefont {Zimdahl}}]{Chimento:2003iea}%
  \BibitemOpen
  \bibfield  {author} {\bibinfo {author} {\bibfnamefont {L.~P.}\ \bibnamefont
  {Chimento}}, \bibinfo {author} {\bibfnamefont {A.~S.}\ \bibnamefont
  {Jakubi}}, \bibinfo {author} {\bibfnamefont {D.}~\bibnamefont {Pavon}}, \
  and\ \bibinfo {author} {\bibfnamefont {W.}~\bibnamefont {Zimdahl}},\ }\href
  {\doibase 10.1103/PhysRevD.67.083513} {\bibfield  {journal} {\bibinfo
  {journal} {Phys.Rev.}\ }\textbf {\bibinfo {volume} {D67}},\ \bibinfo {pages}
  {083513} (\bibinfo {year} {2003})},\ \Eprint
  {http://arxiv.org/abs/astro-ph/0303145} {arXiv:astro-ph/0303145 [astro-ph]}
  \BibitemShut {NoStop}%
\bibitem [{\citenamefont {Copeland}\ \emph {et~al.}(1998)\citenamefont
  {Copeland}, \citenamefont {Liddle},\ and\ \citenamefont
  {Wands}}]{copeland1998}%
  \BibitemOpen
  \bibfield  {author} {\bibinfo {author} {\bibfnamefont {E.~J.}\ \bibnamefont
  {Copeland}}, \bibinfo {author} {\bibfnamefont {A.~R.}\ \bibnamefont
  {Liddle}}, \ and\ \bibinfo {author} {\bibfnamefont {D.}~\bibnamefont
  {Wands}},\ }\href {\doibase 10.1103/PhysRevD.57.4686} {\bibfield  {journal}
  {\bibinfo  {journal} {Phys.Rev.}\ }\textbf {\bibinfo {volume} {D57}},\
  \bibinfo {pages} {4686} (\bibinfo {year} {1998})},\ \Eprint
  {http://arxiv.org/abs/gr-qc/9711068} {arXiv:gr-qc/9711068 [gr-qc]}
  \BibitemShut {NoStop}%
\bibitem [{\citenamefont {Ng}\ \emph {et~al.}(2001)\citenamefont {Ng},
  \citenamefont {Nunes},\ and\ \citenamefont {Rosati}}]{ng2001}%
  \BibitemOpen
  \bibfield  {author} {\bibinfo {author} {\bibfnamefont {S.~C.~C.}\
  \bibnamefont {Ng}}, \bibinfo {author} {\bibfnamefont {N.~J.}\ \bibnamefont
  {Nunes}}, \ and\ \bibinfo {author} {\bibfnamefont {F.}~\bibnamefont
  {Rosati}},\ }\href {\doibase 10.1103/PhysRevD.64.083510} {\bibfield
  {journal} {\bibinfo  {journal} {Phys.Rev.}\ }\textbf {\bibinfo {volume}
  {D64}},\ \bibinfo {pages} {083510} (\bibinfo {year} {2001})},\ \Eprint
  {http://arxiv.org/abs/astro-ph/0107321} {arXiv:astro-ph/0107321 [astro-ph]}
  \BibitemShut {NoStop}%
\bibitem [{\citenamefont {Copeland}\ \emph {et~al.}(2005)\citenamefont
  {Copeland}, \citenamefont {Garousi}, \citenamefont {Sami},\ and\
  \citenamefont {Tsujikawa}}]{Copeland:2004hq}%
  \BibitemOpen
  \bibfield  {author} {\bibinfo {author} {\bibfnamefont {E.~J.}\ \bibnamefont
  {Copeland}}, \bibinfo {author} {\bibfnamefont {M.~R.}\ \bibnamefont
  {Garousi}}, \bibinfo {author} {\bibfnamefont {M.}~\bibnamefont {Sami}}, \
  and\ \bibinfo {author} {\bibfnamefont {S.}~\bibnamefont {Tsujikawa}},\ }\href
  {\doibase 10.1103/PhysRevD.71.043003} {\bibfield  {journal} {\bibinfo
  {journal} {Phys.Rev.}\ }\textbf {\bibinfo {volume} {D71}},\ \bibinfo {pages}
  {043003} (\bibinfo {year} {2005})},\ \Eprint
  {http://arxiv.org/abs/hep-th/0411192} {arXiv:hep-th/0411192 [hep-th]}
  \BibitemShut {NoStop}%
\bibitem [{\citenamefont {Zhai}\ and\ \citenamefont {Zhao}(2005)}]{Zhai2005}%
  \BibitemOpen
  \bibfield  {author} {\bibinfo {author} {\bibfnamefont {X.-H.}\ \bibnamefont
  {Zhai}}\ and\ \bibinfo {author} {\bibfnamefont {Y.-B.}\ \bibnamefont
  {Zhao}},\ }\href {\doibase 10.1393/ncb/i2005-10154-8} {\bibfield  {journal}
  {\bibinfo  {journal} {Nuovo Cim.}\ }\textbf {\bibinfo {volume} {B120}},\
  \bibinfo {pages} {1007} (\bibinfo {year} {2005})},\ \Eprint
  {http://arxiv.org/abs/gr-qc/0508069} {arXiv:gr-qc/0508069 [gr-qc]}
  \BibitemShut {NoStop}%
\bibitem [{\citenamefont {De-Santiago}\ \emph {et~al.}(2013)\citenamefont
  {De-Santiago}, \citenamefont {Cervantes-Cota},\ and\ \citenamefont
  {Wands}}]{DeSantiago:2012nk}%
  \BibitemOpen
  \bibfield  {author} {\bibinfo {author} {\bibfnamefont {J.}~\bibnamefont
  {De-Santiago}}, \bibinfo {author} {\bibfnamefont {J.~L.}\ \bibnamefont
  {Cervantes-Cota}}, \ and\ \bibinfo {author} {\bibfnamefont {D.}~\bibnamefont
  {Wands}},\ }\href {\doibase 10.1103/PhysRevD.87.023502} {\bibfield  {journal}
  {\bibinfo  {journal} {Phys. Rev.}\ }\textbf {\bibinfo {volume} {D87}},\
  \bibinfo {pages} {023502} (\bibinfo {year} {2013})},\ \Eprint
  {http://arxiv.org/abs/1204.3631} {arXiv:1204.3631 [gr-qc]} \BibitemShut
  {NoStop}%
\bibitem [{\citenamefont {Dutta}\ \emph {et~al.}(2016)\citenamefont {Dutta},
  \citenamefont {Khyllep},\ and\ \citenamefont {Tamanini}}]{Dutta:2016bbs}%
  \BibitemOpen
  \bibfield  {author} {\bibinfo {author} {\bibfnamefont {J.}~\bibnamefont
  {Dutta}}, \bibinfo {author} {\bibfnamefont {W.}~\bibnamefont {Khyllep}}, \
  and\ \bibinfo {author} {\bibfnamefont {N.}~\bibnamefont {Tamanini}},\ }\href
  {\doibase 10.1103/PhysRevD.93.063004} {\bibfield  {journal} {\bibinfo
  {journal} {Phys. Rev.}\ }\textbf {\bibinfo {volume} {D93}},\ \bibinfo {pages}
  {063004} (\bibinfo {year} {2016})},\ \Eprint
  {http://arxiv.org/abs/1602.06113} {arXiv:1602.06113 [gr-qc]} \BibitemShut
  {NoStop}%
\bibitem [{\citenamefont {Tsujikawa}(2006)}]{TsujikawaGeneral}%
  \BibitemOpen
  \bibfield  {author} {\bibinfo {author} {\bibfnamefont {S.}~\bibnamefont
  {Tsujikawa}},\ }\href {\doibase 10.1103/PhysRevD.73.103504} {\bibfield
  {journal} {\bibinfo  {journal} {Phys.Rev.}\ }\textbf {\bibinfo {volume}
  {D73}},\ \bibinfo {pages} {103504} (\bibinfo {year} {2006})},\ \Eprint
  {http://arxiv.org/abs/hep-th/0601178} {arXiv:hep-th/0601178 [hep-th]}
  \BibitemShut {NoStop}%
\bibitem [{\citenamefont {Amendola}\ \emph {et~al.}(2006)\citenamefont
  {Amendola}, \citenamefont {Quartin}, \citenamefont {Tsujikawa},\ and\
  \citenamefont {Waga}}]{amendola2006challenges}%
  \BibitemOpen
  \bibfield  {author} {\bibinfo {author} {\bibfnamefont {L.}~\bibnamefont
  {Amendola}}, \bibinfo {author} {\bibfnamefont {M.}~\bibnamefont {Quartin}},
  \bibinfo {author} {\bibfnamefont {S.}~\bibnamefont {Tsujikawa}}, \ and\
  \bibinfo {author} {\bibfnamefont {I.}~\bibnamefont {Waga}},\ }\href {\doibase
  10.1103/PhysRevD.74.023525} {\bibfield  {journal} {\bibinfo  {journal}
  {Phys.Rev.}\ }\textbf {\bibinfo {volume} {D74}},\ \bibinfo {pages} {023525}
  (\bibinfo {year} {2006})},\ \Eprint {http://arxiv.org/abs/astro-ph/0605488}
  {arXiv:astro-ph/0605488 [astro-ph]} \BibitemShut {NoStop}%
\bibitem [{\citenamefont {Chen}\ \emph {et~al.}(2009)\citenamefont {Chen},
  \citenamefont {Gong},\ and\ \citenamefont {Saridakis}}]{ChenPhantom}%
  \BibitemOpen
  \bibfield  {author} {\bibinfo {author} {\bibfnamefont {X.-M.}\ \bibnamefont
  {Chen}}, \bibinfo {author} {\bibfnamefont {Y.-G.}\ \bibnamefont {Gong}}, \
  and\ \bibinfo {author} {\bibfnamefont {E.~N.}\ \bibnamefont {Saridakis}},\
  }\href {\doibase 10.1088/1475-7516/2009/04/001} {\bibfield  {journal}
  {\bibinfo  {journal} {JCAP}\ }\textbf {\bibinfo {volume} {0904}},\ \bibinfo
  {pages} {001} (\bibinfo {year} {2009})},\ \Eprint
  {http://arxiv.org/abs/0812.1117} {arXiv:0812.1117 [gr-qc]} \BibitemShut
  {NoStop}%
\bibitem [{\citenamefont {Landim}(2015)}]{Landim:2015poa}%
  \BibitemOpen
  \bibfield  {author} {\bibinfo {author} {\bibfnamefont {R.~C.~G.}\
  \bibnamefont {Landim}},\ }\href {\doibase 10.1142/S0218271815500856}
  {\bibfield  {journal} {\bibinfo  {journal} {Int. J. Mod. Phys.}\ }\textbf
  {\bibinfo {volume} {D24}},\ \bibinfo {pages} {1550085} (\bibinfo {year}
  {2015})},\ \Eprint {http://arxiv.org/abs/1505.03243} {arXiv:1505.03243
  [hep-th]} \BibitemShut {NoStop}%
\bibitem [{\citenamefont {Landim}(2016{\natexlab{c}})}]{Landim:2015uda}%
  \BibitemOpen
  \bibfield  {author} {\bibinfo {author} {\bibfnamefont {R.~C.~G.}\
  \bibnamefont {Landim}},\ }\href {\doibase 10.1140/epjc/s10052-016-3894-2}
  {\bibfield  {journal} {\bibinfo  {journal} {Eur. Phys. J.}\ }\textbf
  {\bibinfo {volume} {C76}},\ \bibinfo {pages} {31} (\bibinfo {year}
  {2016}{\natexlab{c}})},\ \Eprint {http://arxiv.org/abs/1507.00902}
  {arXiv:1507.00902 [gr-qc]} \BibitemShut {NoStop}%
\bibitem [{\citenamefont {Mahata}\ and\ \citenamefont
  {Chakraborty}(2015)}]{Mahata:2015lja}%
  \BibitemOpen
  \bibfield  {author} {\bibinfo {author} {\bibfnamefont {N.}~\bibnamefont
  {Mahata}}\ and\ \bibinfo {author} {\bibfnamefont {S.}~\bibnamefont
  {Chakraborty}},\ }\href {\doibase 10.1142/S0217732315500091} {\bibfield
  {journal} {\bibinfo  {journal} {Mod. Phys. Lett .A}\ }\textbf {\bibinfo
  {volume} {30}},\ \bibinfo {pages} {1550009} (\bibinfo {year} {2015})},\
  \Eprint {http://arxiv.org/abs/1501.04441} {arXiv:1501.04441 [gr-qc]}
  \BibitemShut {NoStop}%
\end{thebibliography}%

\end{document}